\documentclass{article}
\usepackage{graphicx}
\begin{document}
\fontsize{14}{18pt} \selectfont
\newcommand{\bea}{\begin{eqnarray}}
\newcommand{\eea}{\end{eqnarray}}

\title{Role of quantum fluctuations at high energies}

\author{S.G. Rubin$^{a}$\footnote{e-mail: serg.rubin@mtu-net.ru} ,
H. Kr\"{o}ger$^{b}$\footnote{hkroger@phy.ulaval.ca} , G.
Melkonian$^{b}$\footnote{gmelkony@phy.ulaval.ca}  \\
 $^a$Centre for CosmoParticle Physics ''Cosmion'', \\Moscow
Engineering Physics Institute \\ $^{b}$D\'{e}partement de
Physique, Universit\'{e} Laval, Qu\'{e}bec, Qu\'{e}bec G1K 7P4,
Canada }

\maketitle

\abstract{ In this report we investigate an influence of virtual
particles on the classical motion of a system in
Minkowski and Euclidean spaces. Our
results indicate that fluctuations of fields different from the
main field decelerate significantly its motion at high energies.
The formalism described here is applicable to the majority of
inflationary models. A smallness of temperature fluctuation of
cosmic relic background could be explained within this framework.

We show also that the quantum fluctuations suppress false vacuum
decay in first-order phase transitions at high energies. }

\bigskip
{\Large 1. Massive fields and inflation}

The first models of inflation solved many conceptual problems. The
price was, in general, very small coupling constants. For example,
one of the most promising model, chaotic inflation \cite{Linde90},
with the potential of the form $\lambda \varphi ^4$ gives $\lambda \sim
10^{-13}$ to be compatible with large-scale temperature
fluctuations. To overcome this difficulty, it seems necessary to
take a second field into consideration with a specific coupling
to the first one, as was done in many modern inflationary
scenarios. For example, hybrid inflation \cite{Linde91a} deals with
two coupled scalar fields with connected classical equations of
motion. There are models which take into consideration an
interaction of a classical field with particles, which are produced by
the main field during a classical motion
\cite{Dolgov98,Berera00,Dymnikova00}.

Here we consider only a pure quantum effect of interaction of the
main field with virtual particles of some other sort. Temperature
effects are not considered here. It is known that a cloud of
virtual particles, having inertia, is able to decelerate the motion
of the classical field, which produces this virtual cloud. This
phenomena is well known in solid state physics (polaron effect, see
e.g.\cite{Devreese72,Ru4}).

To show and discuss the main effect, consider only one auxiliary
scalar field $\chi$ as a representative of all other fields coupled
to an inflaton field $\varphi$. The form of interaction is chosen in
such a way to obtain analytical results:
\begin{equation}\label{mi1}
S=\int d^4x\sqrt{-g}\left[ \frac 12\varphi _{,\mu } \varphi ^{,\mu
}-V(\varphi )+\frac 12\chi _{,\mu }\chi ^{,\mu }- \frac
12m_{\chi}^2\chi ^2-\kappa \chi u(\varphi )\right] .
\end{equation}
Here $u(\varphi)$ is the polynomial of $\varphi$ with a power less than
3 for renormalazable theories. The first power of field $\chi$
permits one to obtain final formulae which are valid for some arbitrary
coupling constant $\kappa$. $\chi$ - particles are supposed to be
virtual, and thus the transition amplitude has the form
\begin{equation}\label{mi2}
A(\varphi _i,\chi _i=0\chi ;\varphi _f,\chi _f=0)= \int_{\varphi
_i}^{\varphi _f}D\varphi \int_0^0D\chi \exp \left[ iS\right] .
\end{equation}
Here the field $\chi$ is considered to be rather massive, so that
its classical component is placed at a minimum of its potential.
Integrating out the field variable $\chi$ one arrives at the
effective action
\bea\label{mi3} S_{eff}=\int d^4x\sqrt{-g}\left[ \frac 12\varphi
_{,\mu } \varphi ^{,\mu }-V(\varphi )\right] + \frac{\kappa
^2}2\int d^4x\int d^4x^{\prime }\sqrt{-g}u(\varphi (x))
G(x,x^{\prime })u(\varphi (x^{\prime })) \eea
This expression is exact, but the nonlocal term prohibits to make
analytical predictions. To proceed, let us expand the nonlocal term
(\ref{mi3}) into a power series of $x-x'$ just like in the method
of effective action \cite{Itzykson}. The simplest way is to use the
equation for the Green function \cite{Birrell} in the form
\begin{equation}\label{mi4}
G(x,x^{\prime })=\frac 1{m_{\chi}^2 \sqrt{-g}}\delta (x-x^{\prime
})- \frac 1{m_{\chi}^2}\frac 1{\sqrt{-g}}\partial _\mu \sqrt{-g}
\partial ^\mu G(x,x^{\prime })
\end{equation}
Substituting it into expression (\ref{mi3}) and neglecting higher
derivatives of field $\varphi$ (recall that we are going to deal
with slow motion), one obtains
\begin{equation}\label{mi5}
S_{eff}=\int d^4x\sqrt{-g}\left[ \frac 12\varphi _{,\mu } \varphi
^{,\mu }-V_{ren}(\varphi )-\frac 1{\sqrt{-g}} \frac{\kappa
^2}{2m_{\chi}^4}u(\varphi )\partial _\mu \sqrt{-g}
\partial ^\mu u(\varphi )\right]
\end{equation}
The first term in Eq.(\ref{mi4}) contributes to renormalization of
initial parameters of the potential. The last term in
Eq.(\ref{mi5}) changes the form of kinetic term. As parameters of
potential will be determined by some physical conditions, we will
concentrate on new terms in the kinetic energy which cannot be
incorporated into parameter renormalization. The importance of this
is discussed in \cite{Vil01} in connection with the cosmological
constant problem.

The classical equation of motion can be written
\begin{equation}\label{mi6}
\partial _\mu \sqrt{-g}\partial ^\mu \varphi +
\sqrt{-g}V_{ren}^{\prime }(\varphi )+{\frac{\alpha ^2}{m_{\chi}^2}}
u_\varphi ^{\prime }(\varphi )\partial _\mu \sqrt{-g}
\partial ^\mu u(\varphi )=0 ,
\end{equation}
where $\alpha \equiv \frac{\kappa}{m_{\chi}}$.

Renormalized parameters contain contributions from interactions
with all fields and the field $\chi$ being among them. The latter
contributes to renormalization of the coupling constant
$\lambda_{ren}$ and can be written explicitly - $\delta
\lambda_{ren}^{(\chi )}=-\frac{\alpha ^2}{2}$. The main shortcoming
of the first model of chaotic inflation with the potential of the
form $\lambda_{ren} \varphi^4$ was a smallness of the coupling
constant $\lambda_{ren} (\sim 10^{-13})$ necessary to avoid
contradiction with observations. It means that all terms in the final
expression for $\lambda_{ren}$ including $\delta
\lambda_{ren}^{(\chi )}$ must be cancelled with a high accuracy.
Below it is shown that renormalization of the kinetic term permits
one to weaken restrictions for parameters of the theory.

The inflaton field is considered homogeneous during the period of
inflation, $\varphi =\varphi (t)$ and Eq.(\ref{mi6}) becomes more
simple. One can neglect terms proportional to $d^2 \varphi /dt^2$
and $(d\varphi /dt)^2$ taking into account slow motion of the field
$\varphi$, which takes place under condition
\begin{equation}\label{mi11a}
\varphi \gg \varphi _c  \equiv \frac{m_{\chi}}{{2\alpha }},
\end{equation}
It leads to a classical equation of the form
\begin{equation}\label{mi9}
{\frac{12H\alpha ^2}{m_{\chi}^2 }}\varphi ^2 \dot \varphi +
V_{ren}'(\varphi ) = 0 .
\end{equation}
On deriving this equation, the usual connection of the scale factor
$a(t)$ and Hubble parameter $H$ - $H=\dot a /a$ was assumed to
hold. Consider as a specific case $u(\varphi )=\varphi ^2$ and
$V(\varphi )=\lambda \varphi ^n$. Then the solution of
Eq.(\ref{mi9}) can be written
\begin{equation}\label{mi12}
\varphi (t)=\left[ \varphi _0^{4-n/2}-t/f_n\right]
^{1/(4-n/2)},f_n= \frac{8\sqrt{6\pi }}{n(4-n/2)}\frac{\kappa
^2}{m_{\chi}^4M_{p}\lambda ^{1/2}}.
\end{equation}
Here the following expression for the Hubble parameter
$H=\sqrt{8\pi V(\varphi )/3}/M_P$ was used. The condition of slow
motion is ${\ddot \varphi }<< 12H\varphi ^2 {\dot \varphi
}{\frac{\alpha^2}{m_{\chi}^2}}$. It is important to mention that
the velocity of motion
$$\dot \varphi ={\frac{m^2}{12\alpha ^2\varphi ^2}}\cdot
\frac{V'}{H},$$
obtained from Eqs. (\ref{mi9}) appears to be much smaller than
the usual value \cite{Linde90},
$$\dot \varphi = \frac{V'}{3H},$$
which justifies our initial assumption. First, a 'superslow' stage of
inflation is finished when $\varphi \sim \varphi _c$. Then the stage of
ordinary inflation takes place until ${\ddot \varphi }<<3H{\dot
\varphi}$.

Let us determine an amplitude of quantum fluctuations during the
'superslow' stage of inflation. Fluctuations of noninteracting
fields were investigated before \cite{Linde90}. At the same
time only estimations of order of magnitude are known for
interacting fields \cite{Dolgov92}. The probability of given
fluctuations developed for a cosmological time  $\sim 1/H$ in a causally
connected region with the size $\sim 1/H$ is not small if the action is not
large as compared with unity, $\Delta S\leq 1,\quad \hbar =1$. The main
contribution gives the last term in Eq.(\ref{mi5}):
$$ 1\sim \Delta S_{eff}\sim \frac{2\kappa ^2}{m_{\chi}^4}\frac
1{H^3} \varphi ^2\frac{\Delta \varphi ^2}{H^{-1}} $$
and hence
\begin{equation}\label{mi13}
\Delta \varphi \sim \frac{m_{\chi}}{\alpha \varphi }H.
\end{equation}
Thus, the amplitude of fluctuations are proven to be small as compared
with those in the case of ordinary inflation $H/2\pi$ or as compared
with the estimation $H/\lambda ^{1/4}$ on the basis of the term $V(\varphi
)=\lambda \varphi ^4$, \cite{Dolgov92}. It should be mentioned
that in terms of the quasifield $\widetilde{\varphi}=\varphi ^2$ the high -
energy behaviour looks just like that for a free massive field.

To proceed let us find the field value $\varphi_U$ when our
Universe was born. It is known that $N_U \simeq 60$ e-folgings
enough to explain the main observational data. Taking into account the
connection $N_U =\int _{\varphi_U}^{\varphi_{final}}Hdt$, we obtain
\begin{equation}\label{mi16}
\varphi _U  \simeq \left( \frac{N_U }{2\pi } \right) ^{1/4} \sqrt
{\frac{M_P m_{\chi }}{\alpha }}.
\end{equation}
One can see that our Universe could be nucleated rather late due to
the effect of virtual particles if $m_{\chi}<<M_P$. It does not seem strange
because the first stage of inflation is characterized by a 'superslow'
motion, and the Universe has enough time for an appropriate expansion.

All discussion above holds if quanta of the auxiliary field $\chi$
are quite heavy, and the field is at a minimum of its potential
during the inflationary stage. It takes place if $H<m$. This inequality
ought to hold at least during the first stage of inflation, when
$\varphi_U \geq \varphi \geq \varphi_c$. Some simple calculations give
the estimates
\begin{equation}\label{mi17}
\begin{array}{l}
m_{\chi} > H(\varphi _U )\quad \to \quad \frac{{\sqrt \lambda
}}{\alpha }\leq 0.1 \\ m_{\chi} > H(\varphi _c )\quad \to \quad
m_{\chi}\leq M_P \frac{{\sqrt \lambda  }}{{\alpha ^2 }} \\
\end{array}
\end{equation}%
which do not seem to be very strong. A lower limit for $m_{\chi}$ is
determined at least by quantum corrections, and we have a plausible
estimation $m_{\chi}\geq \kappa$ or, equivalently, $\alpha <1$.

Let us find a parameter range which does not contradict the
observable temperature fluctuations of cosmic background radiation.
Standard calculations \cite{Bardeen83,Mukhanov92} where the expression
(\ref{mi13}) is taken into account leads to the equality
$$\frac{\delta \rho }\rho =16\sqrt{6\pi } N_U\frac{m_{\chi}}{M_P}
\frac{\sqrt{\lambda}}{\alpha},$$
for the potential $\lambda \varphi ^4$. According to the observational data
\cite{COBE}, $\delta \rho /\rho \approx 6\cdot 10^{-5}$ at the
scale of modern horizon. Thus one obtains only connection of
parameters in our model
\begin{equation}\label{mi18}
\frac{{\sqrt \lambda}}{\alpha}\frac{m_{\chi}}{{M_P }}\sim 10^{-8},
\end{equation}
instead of a strong restriction for the parameter $\lambda$.

As a result, we can conclude that the collective motion of the inflaton
field and virtual particles are able to explain observable
small temperature fluctuations in a natural way.

\bigskip

{\Large 2. Suppression of first-order phase transitions by virtual
particles}

\bigskip

In the following we show that the virtual particles
decreases significantly the probability of vacuum decay even at
zero temperature. It could lead to an alternation of the order of phase
transitions at the early stage of evolution of our Universe.

Let us start with the double well potential of the scalar field with
nondegenerate vacua. Following the logic discussed in the beginning
of this paper, consider an auxiliary field $\chi$ with action
(\ref{mi1}). Phase transitions are investigated usually in Euclidean
space, which means a substitution $t \rightarrow i\tau$ in the
formulae written above.

The calculations similar to those in the first part lead to
an effective Euclidean action for the scalar field $\varphi$
\begin{equation}\label{tu9}
\begin{array}{l}
S_E  = \int {d^4 x} \left[ \frac{1}{2}(\partial \varphi )^2  +
V_{ren}(\varphi ) \right] + \frac{{\alpha ^2 }}{2m_{\chi}^2}\int
{d^4 x}\left[ \frac{{\partial u(\varphi (x))}}{{\partial x_\mu
}}\right]^2.
\end{array}
\end{equation}
(Here and below gravitational effects are omitted).The last term can
be interpreted as influence of the virtual $\chi$ - particles.

Let the field $\varphi$ be placed initially at a metastable minimum of
the potential $V_{ren}$. In this case the decay of the vacuum goes by
nucleation and expanding of bubbles with a true vacuum $\varphi_T$
inside it. The outer space is filled with a metastable phase $\varphi
_F$. This process is described by O(4) - invariant solution
$\varphi_B (r)$ of the classical equation of motion in Euclidean space
with the boundary conditions $\varphi_B (0)=\varphi_T ;\varphi_B
(\infty )=\varphi_F $. The probability of the vacuum decay was obtained
in \cite{Coleman77}
\begin{equation}\label{tu3}
\Gamma /V = \left( {\frac{{S_E (\varphi _B )}}{{2\pi }}} \right)^2
\left| {\frac{{Det'\hat D(\varphi _B )}}{{Det\hat D(\varphi _F )}}}
\right|^{ - 1/2} e^{ - S_E (\varphi _B )},
\end{equation}
where the kernel $K$ of operator $\hat D(\varphi )$ has the form $K
(x,y)\equiv \frac{\delta^{2}S_E (\varphi) }{\delta \varphi
(x)\delta \varphi (y)}$.
The effective action in the exponent is the main factor. On the
other hand, as shown below, the value of the effective action
orders of magnitude increases under the influence of virtual
particles. Let us choose the potential in the form
\begin{equation}
V_{ren}=\frac \lambda 8\left( \varphi ^2-a^2\right) ^2+\frac
\varepsilon {2a}\left( \varphi -a\right) . \label{g3}
\end{equation}
The instanton solution of the Euclidean equation of motion for the field
$\varphi$ may be parameterized in the following way
\begin{equation}
\varphi (x)=\varphi_{B} (r)=A\tanh \left( \frac M 2(r-R)\right) -B,
\label{g4}
\end{equation}
were $r^2\equiv \sum_{\alpha =1}^{4}x_\alpha ^2$. Parameters $R$ è
$M$ are to be determined by minimization of the action (\ref{mi9})
while parameters $A$ è $B$ are chosen to satisfy the boundary
conditions
\begin{equation}\label{g4a}
\begin{array}{l}
\varphi _{B} (r \to \infty ) = \varphi _ F  ; \\ \varphi _{B} (r
\to 0) = \varphi _ T  .
\end{array}
\end{equation}
Higher derivatives are neglected in the expression (\ref{tu9}) of the
action. This approximation is correct if
$\partial_{x}\varphi_{B}/m_{\chi}\varphi_{B}<<1$ ($m_{\chi}$ is the
mass of $\chi$ - particles which are formed in a virtual cloud). On
the other hand, the derivative of the instanton trajectory $\partial_x
\varphi_B$ is of the order of the mass of the main particle
$m_{\varphi}$. Numerical calculations indicate that the transition
from a false vacuum to a true one becomes noticeably wider due to the
influence of virtual particles, i.e. $M<<m_{\varphi}$. Thus our
approximation is
valid at least if $m_{\varphi}/m_{\chi} << 1$. A numerical O(4) -
symmetrical solution of the equation

\begin{figure} \includegraphics[scale=0.65]{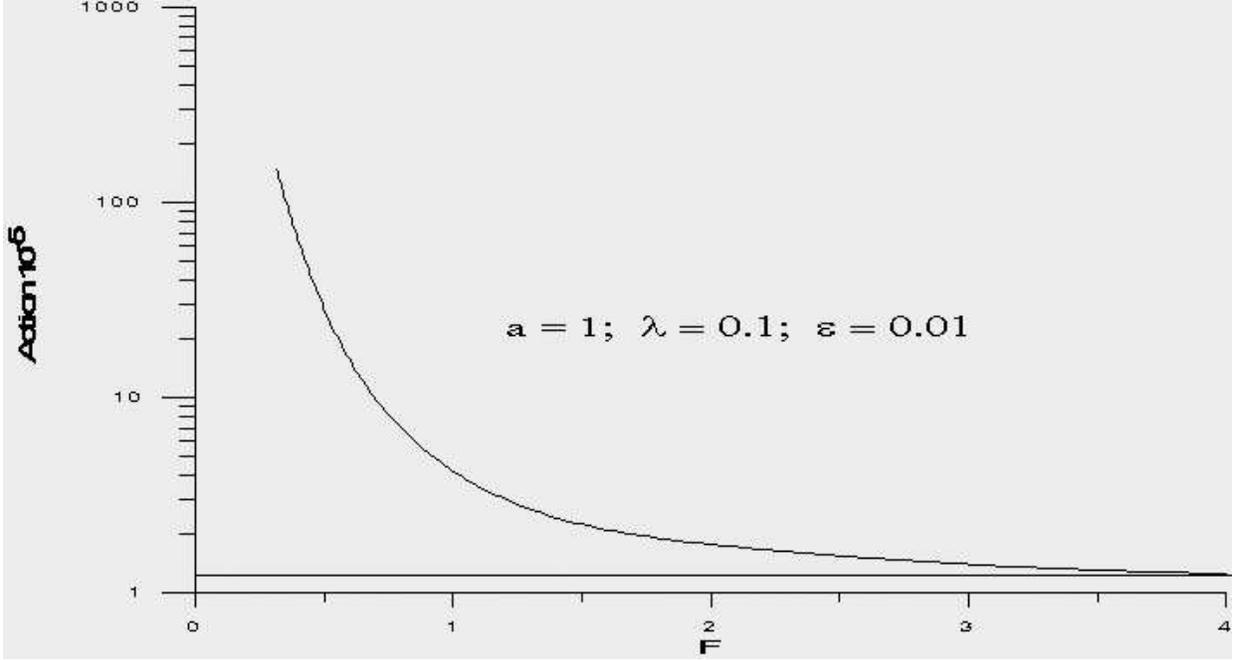}
\caption{Dependence of the effective action on the parameter
$F=2m_{\chi}^2/\alpha^2 $ }. \end{figure}

\begin{equation}\label{g7}
\begin{array}{l}
\partial _r^2 \varphi +\frac{3}{r}\partial _r \varphi -V'(\varphi )+ \\
+ \frac{{\alpha ^2 }}{{m_{\chi}^2 }}u'(\varphi )\left[
{\frac{3}{r}u'(\varphi )\partial _r \varphi  + u''(\varphi
)(\partial _r \varphi )^2  + u'(\varphi )\partial _r^2 \varphi }
\right] = 0
\end{array}
\end{equation}
is supposed to have the form (\ref{g4}). The results of calculations are
represented in Fig.1. The horizontal line marks the standard result. The effect
of virtual particles was not taken into account. Obviously,
the effective action orders of magnitude increases, and the vacuum decay
is exponentially suppressed comparable with the well-known result.

In conclusion, we found that virtual particles at high energies
could significantly influence the classical motion and vacuum
decay. It is shown that in a wide range of parameters the first
stage of inflation consists of a 'superslow' motion of the inflaton
field. One of the useful effect of virtual particles is a
considerable weakening of the conditions on parameters given by
observational data. Tunnelling processes could be strongly
forbidden as compared with a textbook result too.

\bigskip
Acknowledgement

SGR are grateful to A.S. Sakharov and
M.Yu. Khlopov for their interest and useful discussion and A.A.
Starobinsky for critical comments. The work of SGR was partially
performed in the framework of Section "Cosmoparticle physics" of
Russian State Scientific Technological Programme "Astronomy.
Fundamental Space Research", with support of Cosmion-ETHZ and
Epcos-AMS collaborations.


\begin{thebibliography}{10}

\bibitem{Linde90}
A.~D. Linde, {\em The Large-scale Structure of the Universe}
(Harwood Academic
  Publishers, London, 1990).

\bibitem{Linde91a}
A.~D. Linde, Phys. Lett.B {\bf 259},  38  (1991).

\bibitem{Dolgov98}
A.D. Dolgov and S.H. Hansen, hep-ph/9810428  (1998).

\bibitem{Berera00}
A.N. Taylor and A. Berera, astro-ph/0006077  (2000).

\bibitem{Dymnikova00}
I. Dymnikova and M.Yu. Khlopov, Mod.Phys.Lett. {\bf A15},  2305
(2000).

\bibitem{Devreese72}
J.~T. Devreese, {\em Polarons in Ionic Crystals and Polar
Semiconductors}
  (North-Holland, Amsterdam, 1972).

\bibitem{Ru4}
A.~B. Krebs and S.~G. Rubin, Phys.Rev.B {\bf 49},  11808   (1994).

\bibitem{Itzykson}
C. Itzykson and J.-B. Zuber, {\em Quantum Field Theory},   ed.
(McGraw-Hill,
  New York, 1984).

\bibitem{Birrell}
N.D. Birrell and P.C.W. Davies, {\em Quantum Fields in Qurved Space},
Vol.~  of
  {\em  } (Cambridge Univ. Press, Cambridge London New York Sydney,
1982).

\bibitem{Vil01}
J. Garriga and A. Vilenkin, hep-th/0011262, 2001.


\bibitem{Dolgov92}
A.D. Dolgov, Phys. Reports {\bf 222},  309  (1992).

\bibitem{Bardeen83}
J.M. Bardeen, P.J. Steinhardt, and M.S. Turner, Phys.Rev.D {\bf 28},
679  (1983).

\bibitem{Mukhanov92}
V.F. Mukhanov, H.A. Feldman, and R.H. Brandenberger, Phys. Reports
{\bf 215},  203
   (1992).

\bibitem{COBE}
C.L. Bennett et~al., Astrophys.J.Lett. {\bf 464},  1  (1996).

\bibitem{Coleman77}
S. Coleman, Phys. Rev. D {\bf 15},  2929  (1977).

\end{thebibliography}
\end{document}